\documentclass[11 pt,oneside,onecolumn,a4paper]{article}
\usepackage{amsmath}
\usepackage{graphicx}
\DeclareGraphicsExtensions{.eps,.ps,.pdf}
\usepackage{bbm}
\usepackage{bm}% bold math
\usepackage{epsfig}
\usepackage[T1]{fontenc}
\usepackage{esint}
\usepackage{amssymb}

\usepackage{lipsum}
\usepackage{savesym}
mm \leftmargin 5pt \rightmargin 5pt \hoffset= -15mm

\title{Bath energy for correlated initial states
versus information flow in local dephasing channels}
\author{Filippo Giraldi}

\date{\small{School of Chemistry and Physics, University of KwaZulu-Natal\\ 
and National Institute for Theoretical Physics (NITheP)\\
Westville Campus, Durban 4000, South Africa
\vspace{1em}\\Gruppo Nazionale per la
Fisica Matematica (GNFM-INdAM)\\
c/o Istituto Nazionale di Alta Matematica Francesco Severi\\
Citt\'a Universitaria, Piazza Aldo Moro 5, 00185 Roma, Italy}}

\begin{document}

\maketitle

\def\bbm[#1]{\mbox{\boldmath$#1$}}

\vspace{0em}

\vspace{0em}

 \begin{abstract}
Variations of the bath energy are compared with the information flow in local dephasing channels. %in a system of a qubit that interacts locally with a thermal bath. The qubit experiences pure dephasing and no dissipation of energy. 
Special correlated initial conditions are prepared from the thermal equilibrium of the whole system, by performing a selective measurement on the qubit. The spectral densities under study are ohmic-like at low frequencies and include logarithmic perturbations of the power-law profiles. The bath and the correlation energy %, the sum of which is constant, 
alternately increase or decrease, monotonically, over long times, according to the value of the ohmicity parameter, following logarithmic and power laws. Consider initial conditions such that the environment is in a thermal state, factorized from the state of the qubit. In the super-ohmic regime the long-time features of the information flow are transferred to the bath and correlation energy, by changing the initial condition from the factorized to the specially correlated, even with different temperatures. In fact, the low-frequency structures of the spectral density that provide information backflow with the factorized initial condition, induce increasing (decreasing) bath (correlation) energy with the specially correlated initial configuration. By performing the same change of initial conditions, the spectral properties providing information loss, produce decrease (increase) of the bath (correlation) energy.
\end{abstract}

 \maketitle

\section{Introduction}\label{1}
\vspace{-0em}

Nowadays, increasing interest has been devoted to the study of the exchange of energy between an open quantum system and the external environment and to the connections with the flow of quantum information \cite{EnergyFlow1,EnergyFlow2,EnergyFlow3}. 
The analysis performed in Ref. \cite{EnergyFlow1} has shown 
that backflow of energy can be observed in regime of non-Markovianity. Such regime can be interpreted as a flow of quantum information from the environment back in the open system \cite{BnnMarkovPRL2009,nnMarkovNeg2LZPRA2011,FPRA2013,MPRAr2013,MNJP2015}. In local dephasing channels the qubit experiences pure dephasing and no dissipation of energy and the appearance of information backflow is related to the structure of the environmental spectrum \cite{MPRAr2013,MNJP2015,GPRA2017}. In fact, for spectral densities (SDs) that are ohmic-like at low frequencies, information backflow appears uniquely if the ohmicity parameter is larger than a critical value which depends on the temperature of the thermal bath \cite{MPRAr2013}.

In order to find further connections between energy and information flow, here, we consider a qubit that interacts locally with a thermal bath and experiences pure dephasing and no dissipation of energy, for factorized \cite{MPRAr2013} or special correlated initial conditions \cite{M1,M2,M3}. By focusing on the correlated initial configurations, we study how the short- and long-time behaviors of the bath or correlation energy depend on the environmental spectrum. The SDs under study are ohmic-like at low frequencies and include possible logarithmic perturbations of the power-law profiles \cite{GXiv2016,GPRA2017}. For factorized initial states of the qubit and the thermal bath, the directions of the long-time information flow, back in the open system or forth in the external environment, exhibit regular patterns that dependent on the ohmicity parameter \cite{GPRA2017}. In light of these results, we search for connections among variations of the bath or correlation energy, the directions of the information flow and the environmental spectrum in local dephasing channels, for the factorized and  the special correlated initial conditions.

The paper is organized as follows. Section \ref{2} is devoted to the description of the model and the initial conditions. The general class of ohmic-like SDs with logarithmic perturbations is defined in Section \ref{3} The bath and correlation energy and the short- or long-time behavior are analyzed in Section \ref{4} The variations of the bath and correlation energy are compared with the information flow in Section \ref{5} A summary of the result is provided along with conclusions in Section \ref{6} Details of the calculations are given in the Appendix.

\section{Model and initial conditions}\label{2}

The system under study consists in a qubit that interacts locally with a reservoir of field modes \cite{MPRAr2013,MPRA2014,QbtMPRA2014} according to the microscopic Hamiltonian $H$, where $H=H_S+H_{SE}+H_E$. The Hamiltonian of the system, $H_S$, the Hamiltonian of the external environment, $H_E$, and the interaction Hamiltonian, $H_{SE}$, are given by
\begin{equation}
\hspace{-3em}H_S= \omega_0 \sigma_z, \hspace{1em} H_{SE}=\sum_k \sigma_z \left(g_k b_k+g^{\ast}_k b^{\dagger}_k\right), \hspace{1em} H_{E}= \sum_k \omega_k b^{\dagger}_k b_k.
 \label{H}
\end{equation}
In the chosen system of units the Planck and Boltzmann constants are equal to unity, $\hbar=k_B =1$. The transition frequency of the qubit is represented by the parameter $\omega_0$. The rising and lowering operator of the $k$th mode are $b^{\dagger}_k$ and $b_k$, respectively, while $\omega_k$ represents the frequency of the same mode. The coefficient $g_k$ is the coupling strength between the qubit and the $k$th frequency mode. The index $k$ runs over the frequency modes. The $z$-component of the Pauli spin operator \cite{BP,W} is referred as $\sigma_z$. The mixed state of the qubit is described by the reduced density matrix $\rho(t)$, at the time $t$, and is obtained by tracing the density matrix of the whole system, at the time $t$, over the Hilbert space of the external environment \cite{BP}. The model is exactly solvable \cite{RH1,RH2,RH3} and describes a pure dephasing process of the qubit.

Consider initial conditions such that the qubit is decoupled from the external environment, which is represented by a structured reservoir of field modes or by a thermal bath. In the interaction picture, the reduced density matrix describing the qubit, evolves according to the master equation
\begin{equation}
\dot{\rho}(t)=\gamma(t)\left(\sigma_z \rho(t) \sigma_z -\rho(t)\right).  \label{Eq1}
\end{equation}
The function $\gamma(t)$ represents the dephasing rate and depends on the 
SD of the system. At zero temperature, $T=0$, the dephasing rate is labeled here as $\gamma_0(t)$ and reads
\begin{eqnarray}
&&\gamma_0(t)=\int_0^{\infty} \frac{J\left(\omega\right)}{\omega}\,
\sin \left(\omega t \right) \, d\omega.  \label{gamma0} 
\end{eqnarray}
The function $J\left(\omega\right)$ represents the SD of the system and is defined in terms of the coupling constants $g_k$ via the following form,
\begin{equation}
J\left(\omega\right)=\sum_k \left|g_k\right|^2 \delta \left(\omega-\omega_k\right). \label{SD}
\end{equation}
If the external environment is initially in a thermal state, $T>0$, the dephasing rate is represented here as $\gamma_T(t)$ and reads
\begin{equation}
\gamma_T(t)=\int_0^{\infty} \frac{J_T\left(\omega\right)}{\omega}\, d\omega,  \label{gammaT}
\end{equation}
where the effective SD $J_T\left(\omega\right)$ is defined for every non-vanishing temperature $T$ as
\begin{eqnarray}
&&\hspace{-0em}J_T\left(\omega\right)=J\left(\omega\right) \coth \frac{ \omega}{2 T}. \label{JT}
\end{eqnarray}
 For the system under study, the trace distance measure of non-Markovianity \cite{BnnMarkovPRL2009} provides a simple expression of the non-Markovianity measure 
\cite{nnMarkovNeg2LZPRA2011,FPRA2013,MPRAr2013,MNJP2015},
\begin{equation}
\mathcal{N}=\int_{\gamma(t)<0}\left|\gamma(t)\right|e^{-\Xi(t)} d t. \label{N}
\end{equation}
The open dynamics is Markovian if the dephasing rate is non-negative. 
On the contrary, persistent negative values of the dephasing rate witness non-Markovianity and are interpreted as a flow of quantum information from the environment back in the system. Refer to \cite{nnMarkovNeg2LZPRA2011,FPRA2013,MPRAr2013,MNJP2015} for details.

The energy of the open system can be interpreted as the expectation value of the Hamiltonian $H_S$ and is constant, since the Hamiltonian $H_S$ commutes with the total Hamiltonian $H$. The populations of the excited and ground state of the qubit are also constant. The non-equilibrium energy of the bath and the correlation energy can be estimated as the expectation values of the bath Hamiltonian, $H_{E}$, and the interaction Hamiltonian, $H_{SE}$, respectively. The sum of the two energies is referred as the environmental energy and is constant. See Refs. \cite{M1,M2,M3} for details.

\subsection{Special correlated initial states}\label{21}

The analysis performed in Ref. \cite{M2} has shown how the reduced density matrix of the bath, the non-equilibrium energy of the bath, named there as phonon energy, and the non-equilibrium correlation energy evolve in time for special correlated initial states. For the sake of clarity, the main expressions describing such initial conditions, the non-equilibrium energy of the bath and the correlation energy are reported below.

%We focus on initial conditions of the whole system which consist in specially correlated states of the qubit and the thermal bath. 
Following Refs. \cite{M1,M2,M3}, the special correlated initial conditions are prepared from the thermal equilibrium of the whole system at temperature $T$. The system is described by the density matrix $\exp\left(-H/T\right)/Z_0^{\prime}$, where $Z_0^{\prime}$ is a normalization constant, $Z_0^{\prime}=\operatorname{Tr} \left\{\exp
\left(-H/T\right)\right\}$. The symbol $\operatorname{Tr}$ denotes the trace operation over the Hilbert space of the whole system. The qubit is prepared in a pure state $|\phi_0\rangle$ via a selective measurement \cite{measure1,measure2} that induces the whole system in the state $P_0 \exp\left(-H/T\right) P_0/Z_0$. The projector operator $P_0$ represents the effect of the measurement, $P_0=\left|\phi_0\rangle\langle\phi_0\right|$, while $Z_0$ is a normalization constant, $Z_0=\operatorname{Tr} \left\{P_0 \exp\left(-H/T\right) P_0\right\}$. In this way, the whole system is prepared in the initial condition $\rho(0)$, given by $\rho(0)=\left|\phi_0\rangle\langle\phi_0\right|\otimes \rho_E(0)$. The mixed state  of the external environment, $\rho_E(0)$, is given by
\begin{eqnarray}
\rho_E(0)=\frac{\langle \phi_0\left|\exp\left(- H/T\right) \right|\phi_0\rangle}{\operatorname{Tr}_{E}\langle \phi_0\left|\exp\left(- H/T\right) \right|\phi_0\rangle}. \label{rhoE0}
\end{eqnarray}
The symbol $\operatorname{Tr}_{E}$ denotes the trace operation over the Hilbert space of the external environment. The thermal equilibrium of the whole system and the selective measurement on the qubit constrain the thermal bath in the mixed state $\rho_E(0)$. Such state depends on the state $|\phi_0\rangle$ of the qubit and on the interaction Hamiltonian $H_{SE}$. Consequently, in such initial configuration the qubit and thermal bath are correlated. Refer to \cite{M1,M2,M3,LRMP1987,measure3,measure4,measure5,measure6} for details.

\section{ Ohmic-like spectral densities and logarithmic perturbations }\label{3}

The SDs that are usually considered in literature \cite{LRMP1987,W} are ohmic-like at low frequencies and exhibit an exponential cut-off at high frequencies, $J\left(\omega\right)\propto \omega_c\left(\omega/\omega_c\right)^{\alpha_0} \exp \left(-\omega/\omega_c\right)$, are Lorentzian or are characterized by a finite support, $J\left(\omega\right)\propto \omega_c\left(\omega/\omega_c\right)^{\alpha_0}$ for every $\omega \in \left[0,\omega_c\right]$, where $\omega_c<\infty$. The power $\alpha_0$ is referred as the ohmicity parameter, while $\omega_c$ is labeled the cut-off frequency of the reservoir spectrum. The ohmic-like SDs are named sub-ohmic if $0<\alpha_0<1$, ohmic if $\alpha_0=1$ and super-ohmic if $\alpha_0>1$. The description of an experimental setting is beyond the purposes of the present paper. Still, it is worth mentioning that ohmic-like SDs can be engineered in cold environments \cite{SDeng1,SDengBECM2011}. An impurity that is embedded in a double well potential and is immersed in a cold gas reproduces, under suitable conditions, a qubit that interacts with an ohmic-like environment. The ohmicity parameter changes by varying the scattering length and the dimension of the gas, one-, two- or three-dimensional. See Refs. \cite{SDeng1,SDengBECM2011} for details.

 Coherence between the two states of a qubit, backflow of quantum information and the non-equilibrium energy of the reservoir have been largely investigated for ohmic-like SDs \cite{W,BP,RH1,RH2,RH3,MPRAr2013,MNJP2015,M1,M2,M3}. A review of the results is beyond the purposes of the present paper. In light of these studies, one might wonder how the non-Markovian dynamics, the coherence and the non-equilibrium energy of the environment are affected if the SD of the system slightly departs from the physically feasible power-law profiles of the ohmic-like condition at low frequencies. For this reason, two general classes of SDs have been recently considered. In each class, the SDs include the ohmic-like condition, as a particular case, and depart from the low-frequency power laws according to natural or arbitrarily real powers of logarithmic forms \cite{GXiv2016}. The long-time decoherence or recoherence process and the backflow of information are uniquely determined by the ohmicity parameter and, except for special cases, are not affected by the mentioned logarithmic perturbations of the power-law profiles of the ohmic-like SDs \cite{GPRA2017}. In light of these results, we intend to analyze the short- and long-time behavior of the bath energy by considering the ohmic-like and logarithmic-like SDs that are introduced in Refs. \cite{GXiv2016}. For the sake of clarity, we report the definitions and the involved constraints below. For continuous distributions of frequency modes the following constraint on the SD holds \cite{SDP1,ReedSimonBook},
\begin{eqnarray}
\int_0^{\infty} \frac{J\left(\omega\right)}{\omega}\, d\omega<\infty. \label{ConstrSD0}
\end{eqnarray}
The SDs under study are described via the dimensionless auxiliary function $\Omega\left(\nu\right)$ which is defined via the scaling property $\Omega\left(\nu \right)=J \left( \omega_s\nu \right)/ \omega_s $, where $\omega_s$ is a typical scale frequency of the system.

\subsection{First class of spectral densities}\label{SD1}

The first class of SDs is defined by auxiliary functions that are continuous for every $\nu> 0$ and exhibit as $\nu\to 0^+$ the asymptotic behavior \cite{BleisteinBook}
\begin{eqnarray}
&&\hspace{-0em}\Omega\left(\nu\right)\sim 
\sum_{j=0}^{\infty}
\sum_{k=0}^{n_j}c_{j,k} \nu^{\alpha_j}\left(- \ln \nu\right)^k,  \label{o0log} 
\end{eqnarray}
where $\alpha_0>0$, $\infty> n_j\geq 0$, $\alpha_{j+1}>\alpha_j$ for every $j\geq 0$, and $\alpha_j\uparrow +\infty$ as $j\to +\infty$. Since the auxiliary functions $\Omega\left(\nu\right)$ are non-negative, the coefficients $c_{j,k}$ must be chosen accordingly, and the constraint $c_{0,n_0}>0$ is required. The particular case $n_0=0$ provides ohmic-like SDs with ohmicity parameter $\alpha_0$. Consequently, at low frequencies, $\omega\ll \omega_s$, for $n_0=0$ the corresponding SDs are super-ohmic for $\alpha_0>1$, ohmic for $\alpha_0=1$ and sub-ohmic for $0<\alpha_0<1$ \cite{W,LRMP1987}. The logarithmic singularity in $\nu=0$ is removed by defining $\Omega(0)=0$. The above properties and the asymptotic behavior $\Omega\left(\nu\right)= \mathcal{O}\left(\nu^{-1-\chi_0}\right)$ as $\nu\to+\infty$, where $\chi_0>0$, guarantee the required summability of the SDs. Notice that the fundamental constraint (\ref{ConstrSD0}) is fulfilled since the SDs are continuous, $\alpha_0>0$ and $\chi_0>0$. 
Additionally, the Mellin transform $\hat{\Omega}\left(s\right)$ of the auxiliary functions $\Omega\left(\nu\right)$ and the meromorphic continuation \cite{BleisteinBook,Wong-BOOK1989} are required to decay sufficiently fast as $\left|\operatorname{Im} \,s \right|\to+\infty$. Details are provided in the Appendix and in Ref. \cite{GXiv2016}. The definition of this class of SDs is quite simple but the logarithmic powers are restricted to natural values. Arbitrarily positive or negative, or vanishing powers of logarithmic forms are considered in the second class of SDs which is defined below. The insertion of this arbitrariness requires more constraints but allows to perturb the power laws of the ohmic-like profiles with arbitrarily small, positive or negative, powers of logarithmic functions. In fact, arbitrarily small, positive (negative) values of the first logarithmic power $\beta_0$ provide arbitrarily small increases (decreases) in the power-low profiles. In this way, one can evaluate the accuracy of the results that are obtained for the experimentally feasible ohmic-like SDs with respect to logarithmic variations of the low-frequency power-law profiles.

\subsection{Second class of spectral densities}\label{SD2}

The second class of SDs is described by auxiliary functions that exhibit as $\nu\to 0^+$ the asymptotic expansion
\begin{eqnarray}
&&\hspace{-0em}\Omega\left(\nu\right)\sim \sum_{j=0}^{\infty}w_j
\, \nu^{\alpha_j} \left(-\ln \nu\right)^{\beta_j}.  \label{OmegaLog0}
\end{eqnarray}
The powers $\alpha_j$ fulfill the constraints mentioned in Sec. \ref{SD1}, while the powers $\beta_j$ are arbitrarily real, either positive or negative, or vanishing. Since the auxiliary functions $\Omega\left(\nu\right)$ are non-negative, the coefficients $w_j$ must be chosen accordingly, and the constraint $w_0>0$ is required. Again, the logarithmic singularity in $\nu=0$ is removed by setting $\Omega(0)=0$. The auxiliary functions $\Omega\left(\nu\right)$ are required to be continuous and differentiable in the support and summable. %,$\int_0^{\infty}\Omega\left(\nu\right) d \nu<\infty$. 
Again, the summability condition and the choice $\alpha_0>0$ guarantee that the fundamental constraint (\ref{ConstrSD0}) holds. Let $\bar{n}$ be the least natural number such that $\bar{n}\geq\alpha_{\bar{k}}$, where $\alpha_{\bar{k}}$ is the least of the powers $\alpha_k$ that are larger than or equal to unity. The function $\Omega^{\left(\bar{n}\right)}\left(\nu\right)$ is defined as the $\bar{n}$th derivative of the auxiliary function and is required to be continuous on the interval $\left(0,\infty\right)$. The integral $\int_0^{\infty}\Omega\left(\nu\right)\exp\left(-\imath \xi \nu\right) d \nu$ has to converge uniformly for all sufficiently large values of the variable $\xi$ and the integral $\int \Omega^{\left(\bar{n} \right)}\left(\nu\right)\exp\left(-\imath \xi \nu\right) d \nu$ must converge at $\nu=+\infty$ uniformly for all sufficiently large values of the variable $\xi$. The auxiliary functions 
are required to be differentiable $k$ times and the corresponding derivatives must fulfill as $\nu\to 0^+$ the asymptotic expansion
$$\Omega^{(k)}\left(\nu\right)\sim \sum_{j=0}^{\infty}w_j
\, \frac{d^k}{d\nu^k}\left(\nu^{\alpha_j} \left(-\ln \nu\right)^{\beta_j}\right),$$
for every $k=0,1, \ldots,\bar{n} $, where $\bar{n}$ is the non-vanishing natural number defined above. Furthermore, for every $k=0, \ldots,\bar{n}-1$, the function $\Omega^{(k)}\left(\nu\right)$ must vanish in the limit $\nu\to +\infty$. The above constraints are based on the asymptotic analysis performed in Ref. \cite{WangLinJMAA1978}. Notice that in both the classes of SDs under study the auxiliary functions $\Omega\left(\nu\right)$ are non-negative, bounded and summable, due to physical grounds, and, apart from the above constraints, arbitrarily tailored.

\section{The bath energy}\label{4}

Following Ref. \cite{M2}, the non-equilibrium energy of the bath, $\epsilon_E(t)$, is evaluated as 
\begin{equation}
\epsilon_E(t)=\sum_k \omega_k n_k(t), \label{EnDef}
\end{equation}
where $n_k(t)=\operatorname{Tr}\left\{\rho(0) b^{\dagger}_k(t)b_k(t)\right\}$. The 
operators $b^{\dagger}_k(t)$ and $b_k(t)$ represent the rising and lowering operators of the $k$th frequency mode in the Heisemberg picture, respectively, at time $t$. The index $k$ runs over the frequency modes. Let the whole system be initially set in the special correlated condition $\rho(0)$, given by $\rho(0)=\left|\phi_0\rangle\langle\phi_0\right|\otimes \rho_E(0)$ and described in Sec. \ref{21} The initial mixed state of the environment, $\rho_E(0)$, is given by Eq. (\ref{rhoE0}). Under such initial condition, the non-equilibrium energy of the bath \cite{M2} is given by 
\begin{eqnarray}
\epsilon_E(t)=\epsilon_E(0)+d_0 \left(\eta_{1}-\Lambda(t)\right). \label{Et}
\end{eqnarray}
If the distribution of frequency modes is discrete, the initial bath energy \cite{M2} is given by 
\begin{eqnarray}
\epsilon_E(0)=\sum_k \frac{\omega_k}{\exp\left( \omega_k/T\right)-1}+ \eta_{1}, \label{E0}
\end{eqnarray}
while, for a continuous distribution of frequency modes, the initial bath energy reads
\begin{eqnarray}
\epsilon_E(0)=\int_0^{\infty} \frac{\omega r\left(\omega\right)}{\exp\left( \omega/T\right)-1}\, d \omega+ \eta_{1}. \label{E0c}
\end{eqnarray}
The function $r\left(\omega\right)$ denotes the density of the modes \cite{BP,W} at frequency $\omega$. The parameter $\eta_{1}$ is the first negative moment of the SD,
$$\eta_{1}=\int_0^{\infty}\frac{J\left(\omega\right)}{\omega} \,d\omega,$$
while the parameter $d_0$ is defined \cite{M2} as 
$$\hspace{-1em}d_0=2 \left(1+ \langle\ \phi_0\left|\sigma_3\right|\phi_0 \rangle
\frac{\sinh\left( \omega_0/T\right)
-\langle\ \phi_0\left|\sigma_3\right|\phi_0 \rangle\cosh\left( \omega_0/T\right)}
{\cosh\left( \omega_0/T\right)-\langle\ \phi_0\left|\sigma_3\right|\phi_0 \rangle\sinh\left( \omega_0/T\right)}\right). \nonumber
$$
%The first negative moment $\eta_{1}$ of the SD is finite (see Sec. \ref{2}). 
The parameter $d_0$ vanishes for $\left|\langle\ \phi_0\left|\sigma_3\right|\phi_0 \rangle\right|=1$, and 
is positive for \\$\left|\langle\ \phi_0\left|\sigma_3\right|\phi_0 \rangle\right|<1$. Consequently, the bath energy $\epsilon_E(t)$ is 
constant, $\epsilon_E(t)=\epsilon_E(0)$, for $\left|\langle\ \phi_0\left|\sigma_3\right|\phi_0 \rangle\right|=1$, while it is time-dependent for $\left|\langle\ \phi_0\left|\sigma_3\right|\phi_0 \rangle\right|<1$. The function $\Lambda(t) $ reads \cite{M2}
\begin{eqnarray}
&&\hspace{-3em}\Lambda(t)=\int_0^{\infty}\frac{J\left(\omega\right)}{\omega}\, \cos\left(\omega t\right)\, d \omega,
\label{Lt}
\end{eqnarray}
and drives the evolution of the bath energy via Eq. (\ref{Et}). This function can be studied in terms of the SD of the system by following the analysis of the bath correlation function that is performed in Ref. \cite{GXiv2016}.

\subsection{Short- and long-time behavior}\label{41}

Following Ref. \cite{M2}, the interaction energy $\epsilon_{SE}(t)$ is defined as the expectation value of the interaction Hamiltonian $H_{SE}$ and reads 
\begin{equation}
\epsilon_{SE}(t)=\operatorname{Tr}\left\{\rho(0)e^{\imath H t}H_{SE}e^{-\imath H t}\right\}. \label{HSEdef}
\end{equation}
The expectation value of the term $\left(H_{SE}+H_{E}\right)$ of the Hamiltonian is constant and is referred as the environmental energy $\epsilon_{\textrm{env}}$. Such energy is given by $\epsilon_{\textrm{env}}=\operatorname{Tr}\left\{\rho(0)\left(H_{SE}+H_{E}\right)\right\}$. Since the environmental energy is constant, the correlation energy can be evaluated from the bath energy as $\epsilon_{SE}(t)=\epsilon_{\textrm{env}}-\epsilon_{E}(t)$.

At this stage, we start our analysis. We study the short- and long-time behavior of the bath and correlation energy. Over short times the bath energy depends on integral and high frequency properties of the SD. If the SD belongs to the first class (Sec. \ref{SD1}) and decays sufficiently fast over high frequencies, $\chi_0>1$, the bath energy increases quadratically in time for $t \ll 1/\omega_s$,
\begin{eqnarray}
\epsilon_E(t)\sim\epsilon_E(0)+l_E t^2, \label{Etshort}
\end{eqnarray}
where $l_E=d_0 \int_0^{\infty} \omega J\left(\omega\right)\,d \omega/2$. If the SD belongs to the second class (Sec. \ref{SD2}) and $\chi_0>3$, the short-time behavior of the bath energy is the same as the one found for the first class, Eq. (\ref{Etshort}). Over long times, $t \gg 1/\omega_s$, the bath energy tends to the asymptotic value $\epsilon_E\left(\infty\right)$, given by
\begin{eqnarray}
\epsilon_E\left(\infty\right)=\epsilon_E(0)+ d_0 \eta_{1}, \label{Einfty}
\end{eqnarray}
while the correlation energy tends to the asymptotic value $\epsilon_{SE}\left(\infty\right)$, given by $\epsilon_{SE}\left(\infty\right)=\epsilon_{\textrm{env}}-\epsilon_{E}\left(\infty\right)$. If the ohmicity parameter is not an odd natural number the bath energy exhibits logarithmic relaxations for $t \gg 1/\omega_s$,
\begin{eqnarray}
\epsilon_E(t)\sim\epsilon_E\left(\infty\right)+u_0 \left(\omega_s t\right)^{-\alpha_0}
\ln^{n_0}\left(\omega_s t\right), \label{Etlong1}
\end{eqnarray}
where $u_0=-\omega_s d_0 c_{0,n_0} \cos \left(\pi \alpha_0/2\right)\Gamma\left(\alpha_0\right)$. The above relaxations turn into dominant inverse power laws for $n_0=0$,
\begin{eqnarray}
\epsilon_E(t)\sim\epsilon_E\left(\infty\right)+u^{\prime}_0 \left(\omega_s t\right)^{-\alpha_0}, \label{Etlong2}
\end{eqnarray}
where $u^{\prime}_0=-\omega_s d_0 c_{0,0} \cos \left(\pi \alpha_0/2\right)\Gamma\left(\alpha_0\right)$. If the ohmicity parameter is an odd natural number, $\alpha_0=1+2 m_0$, where $m_0$ is a natural number, and $n_0$ is a non-vanishing natural number, the bath energy relaxes over long times, $t \gg 1/\omega_s$, as
\begin{eqnarray}
\epsilon_E(t)\sim\epsilon_E\left(\infty\right)+u_1 \left(\omega_s t\right)^{-1-2m_0}
\ln^{n_0-1}\left(\omega_s t\right), \label{Etlong3}
\end{eqnarray}
where $u_1=(-1)^{1+m_0}\pi n_0 \left(2 m_0\right)!\omega_s d_0 c_{0,n_0}/2$. The above relaxations become dominant inverse power laws if $n_0=1$,
\begin{eqnarray}
\epsilon_E(t)\sim\epsilon_E\left(\infty\right)+u^{\prime}_1 \left(\omega_s t\right)^{-1-2m_0}, \label{Etlong4}
\end{eqnarray}
where $u^{\prime}_1=(-1)^{1+m_0}\pi \left(2 m_0\right)!\omega_s d_0 c_{0,1}/2$. Faster relaxations of the bath energy to the asymptotic value $\epsilon_E\left(\infty\right)$ appear if the ohmicity parameter takes odd natural values and if $n_0$ vanishes. Let $k_0$ be the least non-vanishing index such that $\alpha_{k_0}$ is not an odd natural number, or $\alpha_{k_0}=1+ 2m_{k_0}$, where $m_{k_0}$ and $n_{k_0}$ are non-vanishing natural numbers. We consider SDs such that the index $k_0$ exists with the required properties. The long-time behavior of the bath energy is obtained, in the former case, from Eqs. (\ref{Etlong1}) and (\ref{Etlong2}) by substituting the parameter $\alpha_0$ with $\alpha_{k_0}$ and $n_0$ with $n_{k_0}$, and, in the latter case, from Eqs. (\ref{Etlong3}) and (\ref{Etlong4}) by substituting the parameter $m_0$ with $m_{k_0}$ and $n_0$ with $n_{k_0}$.

For the second class of SDs, the bath energy tends over long times, $t \gg 1/\omega_s$, to the asymptotic value $\epsilon_E\left(\infty\right)$ with relaxations that involve arbitrarily positive or negative, or vanishing powers of logarithmic forms,
 \begin{eqnarray}
\epsilon_E(t)\sim\epsilon_E\left(\infty\right)+\left(\omega_s t\right)^{-\alpha_0}
\left(u_2 \ln^{\beta_0}\left(\omega_s t\right)+u^{\prime}_2\ln^{\beta_0-1}
\left(\omega_s t\right)\right), \label{Etlong5}
\end{eqnarray}
where $u^{\prime}_2=w_0 d_0 \omega_s \beta_0 \left(\cos\left(\pi\alpha_0/2\right)\, \Gamma^{(1)}\left( \alpha_0\right)
-\pi \sin\left(\pi\alpha_0/2\right)\, \Gamma\left( \alpha_0\right)/2
\right)$ and $u_2=w_0 u_0/c_{0,n_0}$. If the ohmicity parameter $\alpha_0$ differs from odd natural values, the dominant part of the above asymptotic form is
 \begin{eqnarray}
\epsilon_E(t)\sim\epsilon_E\left(\infty\right)+u_2\left(\omega_s t\right)^{-\alpha_0}
 \ln^{\beta_0}\left(\omega_s t\right), \label{Etlong6}
\end{eqnarray}
and turns into power laws for $\beta_0=0$,
 \begin{eqnarray}
\epsilon_E(t)\sim\epsilon_E\left(\infty\right)+u_2\left(\omega_s t\right)^{-\alpha_0}. \label{Etlong7}
\end{eqnarray}
If the ohmicity parameter is an odd natural number and $\beta_0$ does not vanish, the bath energy tends to the asymptotic value as 
 \begin{eqnarray}
\epsilon_E(t)\sim\epsilon_E\left(\infty\right)+u^{\prime}_2\left(\omega_s t\right)^{-\alpha_0}
\ln^{\beta_0-1}
\left(\omega_s t\right). \label{Etlong7}
\end{eqnarray}
The relaxations become inverse power laws for $\beta_0=1$,
\begin{eqnarray}
\epsilon_E(t)\sim\epsilon_E\left(\infty\right)+u^{\prime}_2\left(\omega_s t\right)^{-\alpha_0}. \label{Etlong8}
\end{eqnarray}
For the second classes of SDs, the long-time relaxations of the bath energy to the asymptotic value exhibit the same dependence on the low-frequency structure as those found for the first class. %Due to the arbitrariness of the logarithmic powers, for the second class of SDs the relaxations are arbitrarily faster (slower) than the inverse power laws for positive (negative), arbitrarily small values of the first logarithmic power $\beta_0$.

\section{Variations of the bath energy versus information flow
}\label{5}
According to the analysis performed in the previous Section, 
the bath (correlation) energy increases (decreases) quadratically over short times, $t \ll 1/\omega_s$, if the SDs decay sufficiently fast over high frequencies, $\chi_0>1$ for the first class, or $\chi_0>3$ for the second class. Over long times, $t \gg 1/\omega_s$, the bath and the correlation energy alternately increase or decrease, monotonically, to the corresponding asymptotic value. The appearance of each of the two regimes depends uniquely on the ohmicity parameter $\alpha_0$ and is independent of the logarithmic factors that possibly tailor the low-frequency structure of the SD, except for special cases that involve odd natural values of the ohmicity parameter. In fact, for $t \gg 1/\omega_s$, the bath (correlation) energy increases up (decreases down) to the asymptotic value for $0<\alpha_0<1$ and $3+4n<\alpha_0<5+4n$, where $n=0,1,2,\ldots$, if the logarithmic power $n_0$ does not vanish. Same long-time increasing (decreasing) behavior appears for every odd natural value of the ohmicity parameter if $n_0$ vanishes and $3+4n<\alpha_{k_0}\leq 5+4n$, where $n=0,1,2,\ldots$. The power $\alpha_{k_0}$ is defined in Sec. \ref{41} Additionally, increase (decrease) of the bath (correlation) energy is obtained for $t \gg 1 /\omega_s$ if $n_0$ does not vanish and $\alpha_0=1+4l_0$, where $l_0$ is natural valued. If the ohmicity parameter differs from the values reported above, the bath (correlation) energy decreases (increases) monotonically for $t \gg 1/\omega_s$ down (up) to the asymptotic value. For the second class of SDs the bath and correlation energy exhibit the same long-time behavior and the same dependence on the low-frequency profile of the SD as those obtained for the first class. Due to the relationship with the low-frequency environmental spectrum, the loss or gain of bath and correlation energy might be controlled and manipulated over long times by preparing the system in the mentioned correlated initial states and by engineering ohmic-like environments.

For the sake of clarity, we remind how the short- and long-time flow of information in local dephasing channels depend on the low- and high-frequency structure of the SD \cite{GPRA2017}. Let the environment be initially in a thermal state and be factorized from the initial state of the qubit. For the first class of SDs, the information is lost in the external environment over short times, $t \ll 1/\omega_s$. Same short-time behavior is obtained for the second class of SDs and $\chi_0>2$. Over long times, $t \gg 1/\omega_s$, the direction of the information flow is determined by the low-frequency structure of the SD via the ohmicity parameter $\alpha_0$. At non-vanishing temperatures, the information flows back in the open system over long times, $t \gg 1/\omega_s$, for $3+4n<\alpha_0<5+4n$, where $n=0,1,2,\ldots$, if the natural logarithmic power $n_0$ does not vanish, $n_0>0$. Information backflow appears also for every odd natural value of the ohmicity parameter that differs from unity, if $3+4n<\alpha_{k_0}\leq 5+4n$, where $n$ takes natural values, and the natural logarithmic power $n_0$ vanishes. Additionally, information backflow is obtained for every odd natural value $\alpha_0=1+4 l_1$, where $l_1$ is a positive natural number, if $n_0>0$. Same long-time behavior and same relationships with the low-frequency structure of the SD are found in the super-ohmic regime, $\alpha_0>1$, for the second class of SDs. Refer to \cite{GPRA2017} for details.

Straight similarities appear by comparing the present analysis of the bath and correlation energy with the behavior of the information flow for the variety of SDs under study. If the whole system is initially prepared in the special correlated states $|\phi_0\rangle\langle \phi_0|\otimes\rho_E(0)$ that are introduced in Sec. \ref{21}, the variations of bath and correlation energy follow the directions of the information flow that are obtained for factorized initial conditions of the qubit and the thermal state of the bath, even if the temperatures of the initial configurations are different. In fact, over short times, if the SDs decay sufficiently fast at high frequencies, the increase (decrease) of the bath (correlation) energy overlaps with the loss of information. Over long times, in the sub-ohmic and ohmic regime, $0<\alpha_0\leq 1$, for the first class of SDs, the information is lost in the environment and the bath (correlation) energy increases (decreases) for both the classes of SDs. Again, the temperatures in the factorized and the special correlated initial conditions can be different. In the super-ohmic regime, $\alpha_0> 1$, perfect accordance is found, for both the classes of SDs, between the backflow of information, obtained for the factorized initial states, and the increase (decrease) of bath (correlation) energy occurring for the special correlated initial conditions. Same relation appears between the decrease (increase) of bath (correlation) energy and the loss of information in the external environment, over long times and in the super-ohmic regime. Again, such correspondence holds even if temperatures of the initial conditions are different. The above connections hold for logarithmic perturbations of the low-frequency ohmic-like profiles of the SDs.

\section{Summary and conclusions}\label{6}

We have considered a qubit that interacts locally with a bosonic bath. Due to the nature of the interaction, the qubit experiences pure dephasing and no dissipation of energy and reproduces a local dephasing channel. The bath and the correlation energy are evaluated via the expectation value of the bath and the interaction Hamiltonian, respectively. The sum of the bath and the correlation energy is interpreted as the environmental energy and is constant \cite{M2}. The whole system is initially prepared in special correlated states of the qubit and the external environment. Such states are obtained from the thermal equilibrium of the whole system by performing a selective measure on the qubit \cite{M1,M2,M3,LRMP1987,measure3,measure4,measure5,measure6}. We have found that, over short times, the bath (correlation) energy increases (decreases) quadratically in time if the SDs decay sufficiently fast at high frequencies. Over long times, the bath (correlation) energy evolves monotonically towards the asymptotic value: it increases (decreases) in the sub-ohmic regime and also for regular intervals of the ohmicity parameter in the super-ohmic regime, and decreases (increases) otherwise. These asymptotic behaviors are not altered by logarithmic perturbations of the low-frequency ohmic-like profiles, except for special conditions that involve odd natural values of the ohmicity parameter. %These considerations suggest that the long-time behavior of the bath and correlation energy might be controlled and manipulated by engineering environments that are ohmic-like at low frequencies.

If the external environment is set in a thermal state and is factorized from the initial state of the qubit, the long-time flow of quantum information exhibits regular patterns that depend on the ohmicity parameter \cite{GPRA2017}. By changing the initial condition from the factorized states to the special correlated states, even at different temperatures, the long-time features of the information flow are transferred to the bath and correlation energy, in the super-ohmic regime. In fact, over long times, for the variety of SDs under study and in the super-ohmic regime, the same low-frequency spectral properties that provide backflow of information for the factorized initial conditions, induce increase (decrease) of bath (correlation) energy if the whole system is initially set in the special correlated states. Similarly, the low-frequency spectral properties that provide long-time loss of information for the factorized initial states, induce decrease (increase) of the bath (correlation) energy, if the system is initially prepared in the special correlated states. These connections hold for ohmic-like environments and are not altered by logarithmic perturbations of the low-frequency ohmic-like profiles of the SDs. Even if no energy flows in the open system, the present analysis shows a straight relation between the information flow and the variations of the bath or correlation energy, over long times, in local dephasing channels, by properly changing the initial conditions. We believe that the present approach is of interest in the context of control and manipulation of the bath energy by engineering the external environment and setting the system in special correlated initial conditions.

\vspace{0 mm}

\appendix\label{A}
\section{Details}

The short- and long-time behavior of the function $\Lambda(t)$, given by Eq. (\ref{Lt}), is analyzed by introducing the function $K\left(\tau\right)$. Such function is defined via the scaling $K\left(\tau\right)=\Lambda\left(\tau/\omega_s\right)/\omega_s $, and results in the form
\begin{equation}
K\left(\tau\right)=\int_0^{\infty}\frac{\Omega\left(\nu\right)}{\nu} \, \cos \left(\nu \tau\right)\, d\nu.\label{K}
\end{equation}
The Mellin transform of the function $K\left(\tau\right)$ is 
\begin{equation}
\hat{K}\left(s\right)=\cos\left(\frac{\pi}{2}\, s\right)\Gamma(s) \hat{\Omega}(-s).\label{Ks}
\end{equation}
The fundamental strip is $0<\operatorname{Re}\,s<\min\left\{1,\alpha_0\right\}$. If $\chi_0>1$ the function $\hat{K}\left(s\right)$ decays \cite{GradRyz}
 in the strip $\max\left\{-4,-1-\chi_0\right\}<\operatorname{Re}\, s<-2$
 as $\hat{K}\left(s\right)= o\left(\left|\operatorname{Im}\, s\right|^{-5/2}\right)$ for $|\operatorname{Im}\, s|\to+\infty$. %The decay of the function $\hat{K}\left(s\right)$ for $|\operatorname{Im} s|\to+\infty$  
Such decay is sufficiently fast and the singularity in $s=-2$ provides the asymptotic expansion (\ref{Etshort}). Same short-time behavior is obtained for the second class of SDs if $\chi_0>3$, by considering the definition of the function $\Lambda(t)$ and performing the time series expansion of the corresponding integrand function.

As far as the long-time behavior of the function $\Lambda(t)$ is concerned, let the strip $\mu_0\leq \mathrm{Re}\,s\leq \delta_0$ exist such that the function $\hat{\Omega}\left(-s\right)$, or the meromorphic continuation, vanishes in the strip for $|\operatorname{Im} \,s|\to+\infty$ as
\begin{eqnarray}
\hspace{-1em}\hat{\Omega}\left(1-s\right)= O\left(\left|\operatorname{Im}\, s\right|^{-\zeta_0}\right), \label{cond0Omegas2}
\end{eqnarray}
where $\zeta_0>1/2+\delta_0$. The parameters $\mu_0$ and $\delta_0$ fulfill the constraints $\mu_0\in \left(0, \min\left\{1,\alpha_0\right\}\right)$ and $\delta_0 \in \left(\alpha_{k_{1}}, \alpha_{k_{2}}\right)$. The parameter $\alpha_{k_{1}}$ is $\alpha_0$ if $\alpha_0$ is not an odd natural number, or if $\alpha_0$ is an odd number and $n_0$ does not vanish;  otherwise $\alpha_{k_{1}}$ coincides with the parameter $\alpha_{k_{0}}$ which is defined in Sec. \ref{4} The index $k_{2}$ is the least natural number that is larger than $k_{1}$ and such that $\alpha_{k_{2}}$ is not an odd natural number, or it is odd and $n_{k_{2}}$ does not vanish. Under the conditions requested above, the singularity of the function $\hat{K}\left(s\right)$ in $s=\alpha_{k_{1}}$ provides Eqs. (\ref{Etlong1})-(\ref{Etlong4}). The asymptotic forms (\ref{Etlong5})-(\ref{Etlong8}) are obtained for the second class of SDs via the analysis performed in Refs. \cite{WangLinJMAA1978,Wong-BOOK1989}. The increasing (decreasing) behavior of the bath energy over long times is obtained for negative (positive) values of the parameters $u_0$, $u^{\prime}_0$, $u_1$, $u^{\prime}_1$, $u_2$, $u^{\prime}_2$. In this way, the conditions on the ohmicity parameter that provide increasing or decreasing behavior of the bath energy for $t \gg 1/\omega_s$ are obtained. This concludes the demonstration of the present results.


\begin{thebibliography}{0}
\bibitem{EnergyFlow1}G. Guarnieri, C. Uchiyama and B. Vacchini, %\emph{Energy backflow and non-Markovian dynamics}, 
Phys. Rev. A {\bf 93}, 012118 %pp:1--10 
(2016).



\bibitem{EnergyFlow2}G. Guarnieri, J. Nokkala, R. Shmidt, S. Maniscalco and B. Vacchini, %\emph{Energy backflow in strongly coupled non-Markovian continuous-variable systems}, 
Phys. Rev. A {\bf 94}, 062101 %pp:1--10 
(2016).

\bibitem{EnergyFlow3}J. Jing, D. Segal, B. Li and L.-A. Wu, %\emph{Transient unidirectional energy flow and diode-like phenomenon induced by non-Markovian environments}, 
Sci. Rep. {\bf 5}, 15332 %pp:1--8 
(2015).

\bibitem{BnnMarkovPRL2009} H.-P. Breuer, E.-M. Laine and J. Piilo, %\emph{Measure for the Degree of Non-Markovian Behavior of Quantum Processes in Open Systems} 
Phys. Rev. Lett. {\bf 103}, 210401 %pp:1--4 
(2009); E.-M. Laine, J. Piilo and H.-P. Breuer, %\emph{Measure for the Non-Markovianity of quantum processes}, 
Phys. Rev. A. {\bf 81}, 062115 %pp:1--8 
(2010).

\bibitem{nnMarkovNeg2LZPRA2011}Z. He, J. Zou, L. Li and B. Shao, %\emph{Effective method of calculating the non-Markovianity $\mathcal{N}$ for single-channel open systems}, 
Phys. Rev. A {\bf 83}, 012108 %pp:1--6 
(2011).

\bibitem{FPRA2013}F.F. Fanchini, G. Karpat, L.K. Castelano and D.Z. Rossatto, %\emph{Probing the degree of non-Markovianity for independent and common environments}, 
Phys. Rev. A {\bf 88}, 012105 %pp:1--10 
(2013).

\bibitem{MPRAr2013}
P. Haikka, T.H. Johnson and S. Maniscalco, %\emph{Non-Markovianity of local dephasing channels and time-invariant discord}, 
Phys. Rev. A {\bf 87}, 010103(R) %pp:1--5 
(2013).

\bibitem{MNJP2015}
C. Addis, F. Ciccarello, M. Cascio, G.M. Palma and S. Maniscalco, %\emph{Dynamical decoupling efficiency versus quantum non-Markovianity}, 
New J. Phys. {\bf 17}, 123004 %pp:1--11 
(2015).


\bibitem{GPRA2017} F. Giraldi, %\emph{Regular patterns in the information flow of local dephasing channels}, 
Phys. Rev. A {\bf 95}, 022109 %pp:1--13 
(2017).


\bibitem{M1} V.G. Morozov, S. Mathey and G. R\"opke, %\emph{Decoherence in an exactly solvable qubit model with initial qubit-environment correlations}, 
Phys. Rev A {\bf 85}, 022101 %pp:1--10 
(2012).

\bibitem{M2} V.V. Ignatyuk and V.G. Morozov, %\emph{Bath dynamics in an exactly solvable qubit model with initial qubit-environment}, 
Condens. Matter Phys. {\bf 16}, 34001 % pp:1--6 
(2013).

\bibitem{M3} V.V. Ignatyuk and V.G. Morozov, %\emph{Enhancement of coherence in qubits due to interaction with the environment}, 
Phys. Rev. A {\bf 91}, 052102 %pp:1--9 
(2015).


\bibitem{GXiv2016}F. Giraldi, arXiv: 1612.03690v1. 

\bibitem{MPRA2014} C. Addis, G. Brebner, P. Haikka and S. Maniscalco,  Phys. Rev. A {\bf 89}  024101 (2014).



\bibitem{QbtMPRA2014}C.Addis, B. Bylicka, D. Chruscinski and S. Maniscalco, Phys. Rev. A {\bf 90} 052103 (2014).

\bibitem{BP} H.-P. Breuer and F. Petruccione, {\it The Theory of Open Quantum Systems}, Oxford University Press, Oxford (2002).



\bibitem{W} U. Weiss, {\it Quantum Dissipative systems},
3rd ed. World Scientific, Singapore (2008).



\bibitem{RH1} J. Luczka, Physica A {\bf 187} 919 (1990).



\bibitem{RH2} G.M. Palma, K.-A. Suominen and A.K. Ekert, Proc. R. Soc. London, Ser. A {\bf 452} (1996) 567.




\bibitem{RH3} J.H. Reina, L. Quiroga and N.F. Johnson,  Phys. Rev. A {\bf 65 } 032306 (2002).


\bibitem{measure1}K. Kraus, {\it States, Effects, and Operations}, Lecture Notes in Physics, Vol. {\bf 190}, Springer, Berlin (1983).

\bibitem{measure2}V.B. Braginsky and F. Ya Khalili, 
{\it Quantum Measurements}, Cambridge University Press, Cambridge (1992).

\bibitem{LRMP1987}A.J. Leggett, S. Chakravarty, A.T. Dorsey, M.P.A. Fisher, A. Garg and W. Zwerger, Rev. Mod. Phys. {\bf 59}, 1 (1987).

\bibitem{measure3}P. Pechukas, %\emph{Reduced dynamics need not be completely positive} 
Phys. Rev. Lett. {\bf 73}, 1060 %--1062 
(1994).
\bibitem{measure4}P. Stelmachovic and V. Buzek, %\emph{Dynamics of open quantum systems initially entangled with environment:Beyond the Kraus representation},
 Phys. Rev. A {64}, 062106 %pp:1--5 
(2001).
\bibitem{measure5}H. Grabert, P. Schramm and G.L. Ingold, %\emph{Quantum Brownian motion: The functional integral approach}, 
Phys. Rep. {\bf 168}, 115 %--207 
(1988).

\bibitem{measure6}L.D. Romero and J.P. Paz, %\emph{Decoherence and initial correlations in quantum Brownian motion}, 
Phys. Rev. A {\bf 55}, 4070 %--4083 
(1997).

\bibitem{SDeng1}M.A. Cirone, G. De Chiara, G. M. Palma, P. Haikka, S. McEndoo and S. Maniscalco, Phys. Rev. A {\bf 84}, 031602 (2011).

\bibitem{SDengBECM2011}P. Haikka, S. McEndoo, G. De Chiara, G. M. Palma, and S. Maniscalco, Phys. Rev. A {\bf 84}, 031602 (2011).





\bibitem{SDP1}M.P. Woods and M.B. Plenio, %\emph{Dynamical error bounds for continuum discretization via Gauss quadrature rules--A Lieb-Roinson bound approach}, 
J. Math. Phys. {\bf 57}, 022105 %pp:1--19 
(2016).


\bibitem{ReedSimonBook} M. Reed and B. Simon, {\it Methods of Modern Mathematical Physics} Vol. 2, Academics Press, Inc. (1975).






\bibitem{BleisteinBook}
N. Bleistein and R.A. Handelsman, {\it Asymptotic expansion of integrals}, Dover Publications, Inc. New York (1975).

\bibitem{Wong-BOOK1989}R. Wong, {\it  Asymptotic approximations of integrals}, Academic Press, Boston (1989).

\bibitem{WangLinJMAA1978} R. Wong and J.F. Lin, J. Math. Anal. Appl. {\bf 64}, 173 %--180 
(1978).






\bibitem{GradRyz}I.S. Gradshteyn and I.M. Ryzhik, \emph{Table of Integrals, Series and Products} edited by A. Jeffrey (Fifth Edition), Academic Press, New York (2000).


\end{thebibliography}
\end{document}